\documentclass[aps,prl,reprint,showpacs,amsmath,superscriptaddress]{revtex4-1}

\usepackage{graphicx}
\usepackage{amsmath}
\usepackage{array}
\usepackage[dvipdfmx,colorlinks,linkcolor = blue,citecolor = blue,urlcolor = blue]{hyperref}

\begin{document}

\title{Efficient Direct Evaporative Cooling in an Atom Chip Magnetic Trap}

\author{Daniel M. Farkas}
\author{Kai M. Hudek}
\affiliation{JILA, University of Colorado and National Institute of Standards and Technology, Boulder, CO 80309–-0440, USA}
\author{Shengwang Du}
\affiliation{Department of Physics, The Hong Kong University of Science and Technology, Clear Water Bay, Kowloon, Hong Kong, China}
\author{Dana Z. Anderson}
\affiliation{JILA, University of Colorado and National Institute of Standards and Technology, Boulder, CO 80309–-0440, USA}
\email{dana@jila.colorado.edu}

\date{\today}

\begin{abstract}
We demonstrate direct evaporative cooling of $^{87}$Rb atoms confined in a dimple trap produced by an atom chip. By changing the two chip currents and two external bias fields, we show theoretically that the trap depth can be lowered in a controlled way with no change in the trap frequencies or the value of the field at the trap center. Experimentally, we maximized the decrease in trap depth by allowing some loosening of the trap. In total, we reduced the trap depth by a factor of 20. The geometric mean of the trap frequencies was reduced by less than a factor of 6. The measured phase space density in the final two stages increased by more than two orders of magnitude, and we estimate an increase of four orders of magnitude over the entire sequence. A subsequent rf evaporative sweep of only a few megahertz produced Bose-Einstein condensates. We also produce condensates in which raising the trap bottom pushes hotter atoms into an rf ``knife" operating at a fixed frequency of 5\,MHz.
\end{abstract}


\maketitle
Forced evaporative cooling is the dominant technique for achieving the low temperatures and high densities needed to produce Bose-Einstein condensates (BECs) in atomic vapors. For magnetically trapped atoms, forced evaporative cooling uses rf-induced spin flips; first suggested by Pritchard~\cite{Pritchard1989}, this approach eventually led to the generation of the first BECs~\cite{Anderson1995,Davis1995,Bradley1995}. In this work, we study an alternative form of cooling called \textit{direct forced evaporation}. Here, the height of the trapping potential is lowered by \textit{directly} reducing the magnetic fields that constitute the trap.  For most BEC experiments, direct evaporation leads to a reduction in trap frequencies that slows down rethermalization and can reduce cooling efficiency to the point that condensation cannot be achieved. While direct forced evaporation has been used to achieve quantum degeneracy in an optical trap~\cite{Barrett2001,Clement2009}, it has never been demonstrated in a magnetic trap.

\begin{figure*}[htb]
\includegraphics[width=6in]{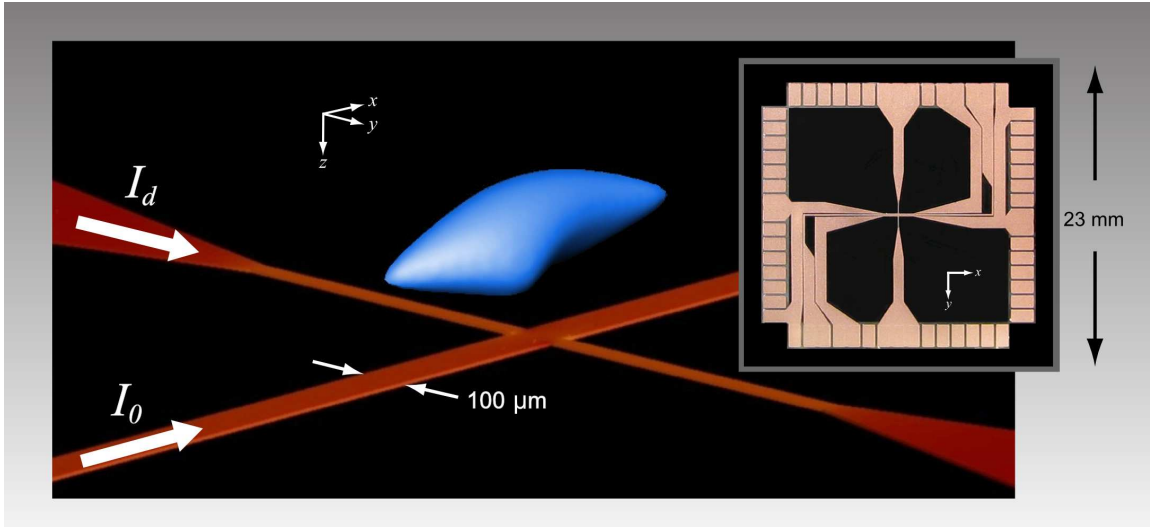}
\caption{A dimple trap is formed underneath the intersection of the 100-micron-wide main trace carrying current $I_0$ and the 25-micron-wide dimple trace carrying current $I_d$. Shown in blue is a 3D contour of the trap at 21\,G, as calculated using the finite-element analysis software LiveAtom, for the parameters listed under the initial stage in Table\,\ref{tab:TrapParameters}. The inset is a photograph of the vacuum side of the 23\,$\times$\,23\,mm atom chip. Of the three traces passing through the center of the chip along the x-direction, only the middle one was used in this work.}
\label{fig:DimpleTrap}
\end{figure*}

In this work, we demonstrate efficient direct evaporative cooling of $^{87}$Rb atoms that are magnetically trapped with an atom chip (see Fig.~\ref{fig:DimpleTrap}). Following an approach first proposed by Du and Oh~\cite{Du2009}, we show theoretically how to directly reduce the chip currents and bias fields of a dimple trap while keeping the trap frequencies and trap bottom constant. Experimentally, we lowered the trap depth by a factor of 20. To maximize the range of cooling, we introduced a six-fold reduction in trap frequency. Even with this loosening of the trap, we estimate that the phase space density (PSD) was increased by four orders of magnitude. At this point, the cloud of atoms was so cold and dense that BECs were produced by raising the trap bottom by a few gauss, pushing the atoms into an rf ``knife" at a single frequency of 5\,MHz.

Our technique dramatically simplifies BEC production by reducing the complexity of the rf system and the power consumption of the atom chip. More specifically, an rf system operating at a fixed frequency can be driven resonantly, greatly reducing the amount of power needed to drive the antenna. Similarly, lowering chip currents reduces chip heating, allowing the atoms to be trapped longer without causing damage from overheating. As such, the techniques and results described here hold promise for ongoing attempts to miniaturize ultracold atom systems for applications out of the laboratory~\cite{Farkas2010,Zoest2010,Bohi2010,Wildermuth2009}, as well as continuous BEC using localized time-independent rf evaporation in a waveguide~\cite{Cren2002,Chikkatur2002,Power2012}.

By trapping atoms only a few hundred microns away from the magnetic field sources, traps with sufficient depth can be generated with only a few amperes of current.  Compared to traps produced with external coils, chip traps can have frequencies that are orders of magnitude higher~\cite{Du2009}. High trap frequencies lead to rapid rethermalization of the atomic cloud during evaporative cooling. Experimentally, several groups have produced BECs with rf evaporation stages as short as 1\,s, more than an order of magnitude faster than experiments using weaker traps~\cite{Farkas2010,Horikoshi2006}.

As shown in Fig.~\ref{fig:DimpleTrap}, the dimple trap is formed from two chip traces that cross each other perpendicularly. Currents $I_0$ and $I_d$ run through the main and dimple traces, respectively. Bias fields $B_{0x}$ and $B_{0y}$ are oriented along the x- and y-directions, respectively. Like most three-dimensional harmonic traps, the dimple trap can be described by eight parameters: the three spatial coordinates of the trap center, the three frequencies of the eigenmodes, the trap depth $U_d$, and the minimum value of magnetic field $B_m$ at the trap center. The dimple trap is always centered below the junction of the two traces, which we take to be at $x=y=0$ in out coordinate system; gravity points in the $-z$ direction. In addition, we assume that the dimple trap has an axisymmetric shape consisting of a single axial eigenmode and two transverse eigenmodes with nearly identical eigenfrequencies.

The trap is now described by five parameters. By leaving the position of the trap center along the z-axis unconstrained, the remaining four trap parameters can be controlled through the four experimental parameters $I_0$, $I_d$, $B_{0x}$ and $B_{0y}$.  This allows the trap depth $U_d$ to be reduced while keeping the trap frequencies and minimum field $B_m$ constant. At the same time, the trap moves closer to the chip.

To mathematically show how direct evaporation can be implemented with a dimple trap, we initially assume $I_d \ll I_0$ and the traces to be infinitely thin. The current $I_0$ and bias field $B_{0y}$ produce an Ioffe-Pritchard trap with the weak axis along the x-axis. The center of the trap is located a distance $z_0$ beneath the origin, where
\begin{eqnarray*}
z_0=\frac{\mu_0}{2\pi}\frac{I_0}{B_{0y}},
\end{eqnarray*}
and $\mu_0$ is the permeability of free space. The gradient $B'_{\perp}$ of the field in the y-z plane is
\begin{eqnarray*}
B'_{\perp}=\frac{B_{0y}}{z_0}=\frac{2\pi}{\mu_0}\frac{B_{0y}^2}{I_0},
\end{eqnarray*}
and the second-order derivative can be expressed
\begin{eqnarray}
B''_{\perp}=\frac{B^{'2}_{\perp}}{B_m}-\frac{B''_{\parallel}}{2}.
\label{eqn:BperpPrimePrime}
\end{eqnarray}
Here, $B''_{\parallel}$ is the second-order field derivative along the x-direction.

We derive $B''_{\parallel}$ by expressing the field in the x-direction as
\begin{eqnarray*}
B_x(x) &=& \frac{\mu_0}{2\pi} \frac{I_d z_0}{z_0^2 + x^2}-B_{0x} \nonumber \\
&\approx& -B_{0x} + \frac{\mu_0}{2\pi}\frac{I_d}{z_0}\left(1-\frac{x^2}{z_0^2}\right).
\end{eqnarray*}
In order to Taylor expand this expression about $x=0$, we assume $x \ll z_0$. At the trap center, we require the minimum value of the field $B_m$ to be greater than 0, or
\begin{eqnarray}
B_m = B_{0x} - \frac{\mu_0}{2\pi}\frac{I_d}{z_0} > 0.
\label{eqn:Bmin}
\end{eqnarray}
The magnitude of the x-component of the total field can be now be expressed
\begin{eqnarray*}
\left|B_x(x)\right| = B_m + \frac{\mu_0}{2\pi}\frac{I_d}{z_0^3}\,x^2,
\end{eqnarray*}
which has a second derivative
\begin{eqnarray}
B''_{\parallel} = \frac{\mu_0}{\pi}\frac{I_d}{z_0^3}.
\label{eqn:BparaPrimePrime}
\end{eqnarray}

The trap depth $U_d$ is equal to the difference between $B_m$ and the smallest potential the atoms must overcome to escape the trap.  Here, the smallest potential is simply the magnitude of the bias field, which yields
\begin{eqnarray}
U_d = \sqrt{B_{0x}^2 + B_{0y}^2} - B_m.
\label{eqn:TrapDepth}
\end{eqnarray}

\begin{table*}[htb]
		\begin{tabular*}{\textwidth}{@{\extracolsep{\fill}} p{2cm}|cccc|cccc|c}
			\hline \hline
            \noalign{\smallskip}
			Stage & $I_0$\,(A) & $I_d$\,(A)& $B_{0x}$\,(G) & $B_{0y}$\,(G)~~~ & $U_d$\,(G) & $\bar{\omega}$\,(Hz) & $B_m$\,(G) & $z_0$\,($\mu$m)~~~ & Duration\,(ms)~~~~\\
			\noalign{\smallskip}
            \hline
            \noalign{\smallskip}
            Initial    & 3.25 & 1.25 & 25.2 & 42.3 & 41.6 &  983 & 7.7 & 137 & N/A  \\
            A          & 2.2  & 1.25 & 14.4 & 19.8 & 21.9 &  536 & 2.5 & 204 & 400 \\
            B          & 1.1  & 0.55 & 8.1  & 8.1  & 7.9  &  216 & 3.6 & 222 & 400 \\
            C          & 0.58 & 0.3  & 4.7  & 4.1  & 3.9  &  144 & 2.3 & 219 & 700 \\
            D          & 0.23 & 0.06 & 3.8  & 3    & 1.9  &  171 & 2.7 & 119 & 400 \\
            Compress   & 3.25 & 1.25 & 14.8 & 36   & 38   & 1652 & 0.5 & 175 & 400 \\
            Final RF   & 3.25 & 1.25 & 20.8 & 36   & 35   &  785 & 6.1 & 165 & 800 \\
            \noalign{\smallskip}
			\hline \hline
		\end{tabular*}
	\caption{Trap parameters used for evaporative cooling. Direct evaporation is performed during stages A through D. The final RF stage only applies to the single-frequency rf evaporation technique described in the text.}
	\label{tab:TrapParameters}
\end{table*}

By solving Eqns.~(\ref{eqn:BperpPrimePrime})~through~(\ref{eqn:TrapDepth}) numerically, the solutions for $I_0$, $I_d$, $B_{0x}$, and $B_{0y}$ were used to numerically calculate trap parameters using a more realistic model that accounts for the 100-micron width of the main trace and the 25-micron width of the dimple trace (we ignore the 10-micron thickness of the traces). Trap parameters calculated with the two models agree with each other to within a few percent provided that the trap position $z_0$ is greater than the trace width. As the trap depth is lowered, the position $z_0$ of the trap center moves toward the chip. When the trap center is closer to the chip than the trace width, the traces are no longer infinitely thin and the simple model derived above is no longer valid.

Due to the parameters of our initial trap and the limitations imposed by the finite width of the chip traces, keeping the trap frequencies constant would have implied a trap depth reduction by only a factor of a few. To obtain greater reductions and ensure that $z_0 \geq$ 100$\,\mu$m, we intentionally loosened a trap that was initially very tight ($\bar{\omega}\approx1$\,kHz, where $\bar{\omega}$ is the geometric mean of the three trap eigenfrequencies). The direct evaporation sequence was divided into four stages labeled A through D, with each stage reducing the trap depth by approximately a factor of two. The final values for the experimental parameters of each stage were optimized for maximum PSD and are shown in Table~\ref{tab:TrapParameters}, along with the corresponding numerically calculated trap parameters. In each stage only linear ramps were used, and the final experimental values of one stage were used as the initial values of the following stage.

The atom chip pictured in Fig.~\ref{fig:DimpleTrap} was anodically bonded to the top of a 23\,$\times$\,23\,mm cross-section glass cell which forms the science chamber of a two-chamber UHV cell. Pressure less than $1\times 10^{-9}$\,torr was maintained with a 2\,l/s ion pump and nonevaporable getters. The UHV cell, optomechanics and lasers, and experimental procedures for cooling, trapping, transporting, chip loading, rf evaporation and absorption imaging are described in detail in Ref.~\cite{Farkas2010}.

\begin{figure}
\includegraphics[height=6in]{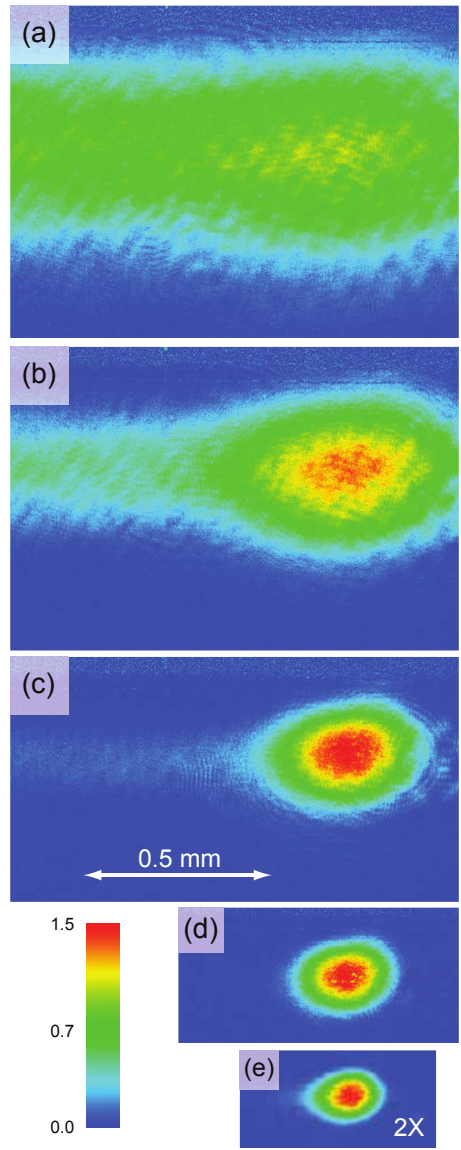}
\caption{Absorption images showing OD after 3\,ms of free expansion. (a) $3.6\times10^7$ atoms in the initial chip trap; (b) $1.8\times10^7$ atoms after direct evaporation stage A; (c) $5.6\times10^6$ atoms after stage B; (d) $2.6\times10^6$ atoms after stage C; and (e) $1.0\times10^6$ atoms after stage D. Here the magnification is doubled.}
\label{fig:ImagesAfterEachStage}
\end{figure}

The images in Fig.~\ref{fig:ImagesAfterEachStage} show the optical depth (OD) of atoms in the initial chip trap and after each of the four direct evaporation stages. Even though both the trap frequencies and total atom number were reduced during stages A and B, the OD has visibly increased, a clear indication that phase space density has increased. Because of the relatively high temperature of the cloud after stages A and B, and the trap's close proximity to the atom chip, we were not able to directly measure the temperature. At the end of stages C and D, the OD remained unchanged even though the number of atoms was roughly halved during each of these stages. For the entire direct evaporation sequence, the trap depth was reduced by a factor of 20 while $\bar{\omega}$ was reduced by less than factor of 6 (Table~\ref{tab:TrapParameters}). In fact, for stages C and D combined, the trap depth was reduced by a factor of 4 with virtually no change in $\bar{\omega}$. The total direct evaporation sequence (stages A through D) took approximately 2\,s.

After stages B, C, and D, the atoms were dropped and allowed to freely expand between 2 and 10\,ms. Each time-of-flight image was fit to a two-dimensional Gaussian function, and the fitted widths were used to extract the in-trap size and temperature of the cloud, shown in Table~\ref{tab:TOFmeasurements}. Using the measured atom number (Fig.~\ref{fig:ImagesAfterEachStage}) and assuming an axisymmetric trap, the peak density was calculated and was used to calculate the PSD shown in Table~\ref{tab:TOFmeasurements}. During stages C and D, the PSD increased by two order of magnitude. Assuming similar efficiencies during stages A and B, we estimate that PSD was increased by almost four orders of magnitude over the entire sequence.

\begin{table}[htb]
		\begin{tabular}{p{0.9cm}|ccccc}
			\hline \hline
            \noalign{\smallskip}
			Stage &~$T_x$\,($\mu$K)~&~$T_y$\,($\mu$K)~&~$\sigma_x$\,($\mu$m)~&~$\sigma_y$\,($\mu$m)~&~PSD~\\
			\noalign{\smallskip}
            \hline
            \noalign{\smallskip}
            B &   94(1) & 50.6(3) & 254(7) & 143(3) & $6.0(5)\times 10^{-7}$ \\
            C & 28.5(2) & 21.8(1) & 160(2) & 86(2)  & $6.7(4)\times 10^{-6}$ \\
            D & 14.0(1) & 10.6(1) & 122(1) & 38(2)  & $5.2(9)\times 10^{-5}$ \\
            \noalign{\smallskip}
			\hline \hline
		\end{tabular}
	\caption{Temperatures $T_x$ and $T_y$, in-trap cloud sizes $\sigma_x$ and $\sigma_y$, and phase space densities PSD at the end of direct evaporation stages B, C, and D.}
	\label{tab:TOFmeasurements}
\end{table}

As another clear indication that PSD increased due to direct evaporation, we produced a BEC by adding a single stage of rf evaporation. Here, the atoms were first adiabatically compressed into a deep, tight trap (Table~\ref{tab:TrapParameters}); a subsequent rf sweep of only a few megahertz was needed to produce a BEC.

\begin{figure}[htbp]
\includegraphics[width=3.25in]{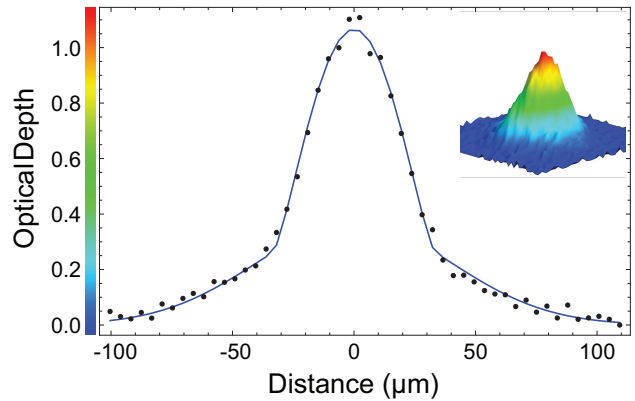}
\caption{The measured OD of a BEC produced using a combination of direct evaporation and single-frequency RF evaporation. The image was taken after a free expansion of 4\,ms. The black data points are an average of the three rows of pixels that pass horizontally through the center of the cloud. The blue line represents the best-fit to the sum of a Gaussian thermal component and paraboloid condensate core. The clear presence of a ``kink" in the wings indicates partial condensation. The cloud contains $21(1)\times10^3$\,atoms with a condensate fraction of 0.24(1).}
\label{fig:BECimage}
\end{figure}

The substantially shorter rf sweep allowed us to demonstrate single-frequency rf evaporation. Here, the rf frequency was fixed at 5\,MHz. Over 800\,ms, the trap bottom was raised by increasing $B_{0x}$, pushing hotter atoms into the rf knife. An OD image of a BEC produced with this method is shown in Fig.~\ref{fig:BECimage}. Taken after 4\,ms of free expansion, there are $21(1)\times10^3$ atoms in the cloud, with a condensate fraction of 0.24(1). The parameters of the final trap are listed in the last line of Table~\ref{tab:TrapParameters}. Note that the calculated value of $B_m$ corresponds to a spin-flip frequency of 4.25\,MHz, slightly less than the 5\,MHz rf drive frequency. We attribute this slight discrepancy to a measured 1\,G stray magnetic field originating from the ion pump's magnets.

When we reduced the trap position $z_0$ to less than 100\,$\mu$m, we observed significant atom loss and loosening of the trap. As described above, trap loosening is due to the fact that the traces appear more like sheets of current rather than infinitely-thin lines as atoms approach the chip surface. The direct evaporation sequence could be further extended by making the traces narrower. However, the trace width is limited by the maximum permissible current density.

Direct evaporative cooling is a useful tool for producing ultracold gases that can complement more traditional rf evaporation techniques. Theoretically we showed that by changing four experimental parameters -- two chip currents and two bias fields -- the trap depth can be reduced with no change in the trap frequencies. We demonstrated direct evaporation on a sample of $^{87}$Rb atoms confined in an atom-chip-based dimple trap. Because of the width of the chip traces, we intentionally introduced a slight weakening of the trap in order to attain shallower trap. Nevertheless, phase space density increased by an estimated four orders of magnitude. Less than 1\,s of subsequent rf evaporation produced a BEC. The ability to produce a BEC using rf evaporation at a fixed frequency can be used to reduce the size and power consumption of the rf system. As such, the techniques demonstrated here could have potential uses for further miniaturizing ultracold-matter apparatus by eliminating the need for a complex rf system and high powers being run through the atom chip.

\begin{acknowledgments}
This work was supported in part by the Defense Advanced Research Projects Agency, the Army Research Office (W911NF-04–1-0043), the Air Force Office of Scientific Research (Grant
No. FA9550-10-1-0135), and the National Science Foundation through a Physics Frontier Center (PHY0551010). S. Du was supported by the Hong Kong Research Grants Council (Project No. HKU8/CRF/11G).
\end{acknowledgments}

%

\end{document}